\documentclass[12pt]{article}
\usepackage{epsfig}

\def\beq{\begin{equation}}
\def\eeq#1{\label{#1}\end{equation}}
\def\eeqn{\end{equation}}

\def\beqa{\begin{eqnarray}}
\def\eeqa#1{\label{#1}\end{eqnarray}}
\def\eeqan{\end{eqnarray}}

\let\bar=\overbar

\def\Dslash{\not{\hbox{\kern-4pt $D$}}}
\def\dslash{\not{\hbox{\kern-2pt $\del$}}}

\def\msb{{\bar{\ssstyle M \kern -1pt S}}}

\def\Title#1{\begin{center} {\Large {\bf #1} } \end{center}}

\begin{document}

\Title{Diquark degrees of freedom in the EOS and the compactness 
of compact stars}

\bigskip\bigskip


\begin{raggedright}  

{\it J. E. Horvath, G. Lugones \index{}\\
Instituto Astron\^omico e Geof\'{\i}sico/Universidade de S\~ao Paulo\\ 
Av. M. St\'efano 4200, Agua Funda, 04301-904 S\~ao Paulo SP, BRAZIL\\}
and
{\it J.A. de Freitas Pacheco\index{}\\
Observatoire de la C\^ote d'Azur\\
BP 4229, F-06304, Nice Cedex 4, FRANCE\\}
\bigskip\bigskip
\end{raggedright}

\section{Introduction} 

"Old" questions of relativistic 
astrophysics, like the internal structure of compact 
stars, have received a lot of attention recently since they are very important for a solid knowledge of the QCD diagram in the low-$T$ , high $\mu$ region as 
discussed along this Workshop. 
Among the most simple forms of investigating (indirectly) the nature of 
high-density matter, more precisely of the equation of state connecting 
the pressure and the energy density; is by obtaining accurate determinations 
of masses and radii. A comparison 
of the static models generated by integration of the 
Tolman-Oppenheimer-Volkoff 
equations with observed data should reveal a great deal of information 
about the equation of state. 

While astronomers and physicists alike have longely dreamed of such 
determinations 
we have only recently achieved sufficient accuracy to perform some key tests 
on selected objects, although it should be acknowledged that three 
decades of compact star astrophysics had produced quite clever arguments to 
determine masses and radii, even in the cases in which they proved to be wrong. 

A paradigmatic example of the above assertions is the celebrated 
determination 
of the binary pulsar mass PSR 1913+16, believed to be accurate to the fourth 
decimal place  \cite{vankerkwijk}. Methods based on combinations of spectroscopic 
and photometric techniques have been recently 
perfectioned, and have confirmed 
that at least one X-ray source is significantly above the "canonical" 
1.44 $M_{\odot}$; namely Vela X-1 for which a value of 
$1.87^{+0.23}_{-0.17} \, M_{\odot}$ has been obtained \cite{vankerkwijk}. 

It is also possible that at least 
some compact stars are extremely compact, or in other words, that their 
radii are $\sim 30-40 \%$ less than the cherished $10 \, km$ for 
$M \, \sim 1 \, M_{\odot}$. Indeed, this is the claim of 
the analysis of  \cite{li} of the binary 
Her X-1 ($M = 0.98 \pm 0.12 \, M_{\odot}$ and $R = 6.7 
\pm 1.2 \, km$) and  \cite{pons} in the case of the isolated nearby RX J1856-37 
($M = 0.9 \pm 0.2 \; M_{\odot}$ and $R = 6^{-1}_{+2} \, km$). In fact, such compactness is 
extremely difficult (impossible?) to model using underlying equations 
of state based on hadrons alone, and a "natural" alternative would be 
to consider deconfined matter. Naturally, these results will be checked and 
scrutinized for confirmation.

Could this be related to high-density deconfined matter ? Perhaps, since 
there is hope to achieve deconfinement densities inside compact stars.
Also, there is some evidence that diquarks (e. g. a spin-0, 
color-antitriplet 
bound state of two quarks) might occur  as a  component in  the QCD 
plasma (see \cite{anselmino} and references therein for a review). 
Such diquarks would be expected to be favoured by Bose statistics 
and they are quit helpful to model the low-energy 
hadronic properties. 
Of course,  at very high densities,  characterized by interquark 
distances less 
than  $(10  GeV^2 )^{-1/2}$,  diquarks  lose  their  identity and 
must eventually dissolve into quarks, even if there is no clear 
consensus about the onset of the asymptotic regime. 

According to this picture we may regard a diquark as any 
system of two quarks considered collectively. Diquark correlations 
arise in part from spin-dependent interactions between two 
quarks. Regardless of the exact mechanism for their origin, it is 
imperative to have some gain of energy (binding) for the stellar models to 
work (see below). The bound state is a quite strong assumption, and is not easy to prove (or disprove) 
in the absence of a reliable way of performing sensible calculations. 
As a working hypothesis (quite analogous to the well known strange matter), 
we shall assume simply that this bound state actually exists.

\section{Bag-inspired quark-diquark Model} 

Some time ago it was the interesting controversy about the compactness of 
Her X-1 that prompted us \cite{horvath1} to consider models in which diquarks 
are considered fundamental digrees of freedom and treated as effective 
bosons. The results were derived using an effective $\lambda \phi^{4}$ 
model and a derived equation of state given in \cite{CSW}, shown to be 
valid in this case. We shall not repeat the analysis here, but just point 
out that (as expected) a self-bound quark-diquark mixture indeed produced 
very compact models (see Fig.1, curve A). In those models there are two 
quantities that had to be treated parametrically: the diquark mass 
$m_{D}$ and the vacuum energy density $B$. There is no relation between the 
two included and therefore other possibilities than the $m_{D} = 575 MeV$ 
$B = 57 MeV fm^{-3}$ could be considered. However, it is desirable to 
reduce the freedom in the parameters to see whether the compactness 
can be altered.
 
\section{QMDD Model}
 
In a more recent attempt we have modeled the quark - diquark  plasma 
as a relativistic gas 
of  $u$ and $d$ quarks, and  $(ud)$ diquarks. We treated non-perturbative 
effects in the quark mass-density-dependent model of confinement (QMDD model) 
extended for the diquarks assuming $m_D = m_{D0} + C/ n_B^{1/3}$, analogously 
to quarks. Some information on 
the parameter $C$  to describe absolutely stable SQM 
has been found in the range (155-171)$^2$  $MeV^2$ \cite{peng2000c61} based on known stability arguments.
The expression for the thermodynamic potential $\Omega$ was derived 
in the thermodynamic limit for the QMDD  model extended 
for bosons, and the usual thermodynamical   quantities   derived from 
$\Omega$ with some care(see \cite{peng2000c61}). 
 
Because $\Omega$ depends on the baryon number $n_B$ through the masses 
$m_i$; the standard relation $P= 
-\Omega$ \cite{peng2000c61,benvenuto95} is not 
valid and we have an extra term generated by this density dependence; namely 
$P = -\frac{ \partial (V \Omega)}{\partial V }  = 
\sum_i \bigg( - \Omega_i + n_B \frac{\partial m_i}{\partial n_B} 
\frac{\partial \Omega_i}{\partial m_i} 
\bigg) $. 
As stressed in previous works \cite{benvenuto95}, the 
term  $n_B   \frac{\partial  m_i}{\partial   n_B}  \frac{\partial 
\Omega_i}{\partial m_i}$ allows the pressure to be zero at finite 
baryon number, playing the role of the Bag Constant B in the  MIT 
Bag Model. Needless to say, this is a property related 
to the parametrization of the mass only, and does not depend on the 
particle statistics of the particles. 
However, in the case of bosons this extra term does not contribute at 
zero temperature. 
As in the standard case of massive bosons  at $T=0$ 
the energy density reduces to $E = \sum_i n_i m_i$. 
Diquarks do not contribute to the pressure as they are all in the ground 
state, but do contribute to the energy density, the equation of state 
is expected to be quite soft. 

These equations have been supplemented with the conditions  
of equilibrium between $D$, $u$  and $d$. By imposing very general stability 
and equilibrium 
conditions we found an absolute lower limit on 
$m_{D0}$ $m_{D0,lim} =  \frac{C}{n_b^{1/3}} 
= 230 \bigg( \frac{n_0}{n_B} \bigg)^{1/3} MeV$. Electrical charge neutrality and baryon number conservation were enforced to calculate the equation of state for each $n_B$ once the 
parameters $(C, m_{D0})$ were given. 
The numerical results were scanned for stability of the mixture . 
Stability (and hence self-binding \cite{foton}) 
occurs inside a pretty large region in the ($C^{1/2},M_{D0}$) plane. 
This region of absolute 
stability allows the existence of pure quark-diquark stars.It is curious 
to remark that the {\it same} parametrization used for simulating the confinement 
produces a behaviour of the diquark mass that mimics the uprising part 
of a gap curve (i.e. increasing binding) and thus energetically favoured 
diquarks as the density increases. This is {\it not} inconsistent with the 
idea of dissociation of the diquarks at still higher densities, but in any 
case it is not expected that a mass density-dependent model can solve the 
complicated structure of a theory at higher densities (see \cite{german} for 
a through discussion of this and other details of the model).

\section{Discussion}

Three examples of the QMDD diquark model are shown in Fig. 1. The 
resulting equations of state are very  soft (models B and C), as expected, but fall
short to explain the claimed compactness by at least $\Delta R \sim 1 \, km$, even  
in the most favourable case (model B). When compared  with the equation of 
state for quark-diquark 
matter presented in \cite{horvath1} (model A) and the softest models of MIT 
bag models \cite{MIT} (model D) we may assert that the stellar sequences 
are in between the former and the latter. Diquarks may be 
relevant for compact star structure 
because they allow the existence of stable compact stars with masses of 
$\sim  1 M_{\odot}$ and very small radii. 
For comparison, the inferred radius and mass of both Her X-1 and RX J1856-37 
are shown with 
their respective error bars. Li {\it et al.} models SS1 and SS2 
\cite{ignazio}(not shown in the figure) are examples of 
other self-bound 
models of the equation of state which do produce very compact stars, 
in fact also capable of matching the values of Her X-1  claimed in 
\cite{li} and of RX J1856-37  \cite{pons}. We may state that if evidence 
for extreme compactness holds, diquarks may be relevant  
for model building . 
\begin{figure}[htb]
\begin{center}
\epsfig{file=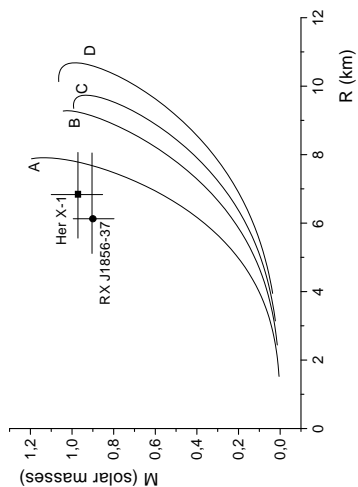,height=2.5in}
\caption{The mass-radius plane, stellar sequences labeled as indicated in the text}
\label{fig:bohr}
\end{center}
\end{figure}

\bigskip
G. Lugones  acknowledges the IAG-S\~ao Paulo for hospitality 
and the financial support received from  the 
Funda\c c\~ao de Amparo \`a Pesquisa do Estado de 
S\~ao Paulo. J.E. Horvath wishes to acknowledge 
the CNPq Agency (Brazil) for partial financial support. The Scientific
Organizers of the meeting and Mrs. Ana Rey are acknowledged for 
making JEH stay in Copenhaguen a very pleasant experience and 
partial financial support to attend the event. I. Bombaci, S. Balberg and 
S. Fredricksson have shared with him 
several interesting discussions.

\def\Discussion{
\setlength{\parskip}{0.3cm}\setlength{\parindent}{0.0cm}
     \bigskip\bigskip      {\Large {\bf Discussion}} \bigskip}
\def\speaker#1{{\bf #1:}\ }
\def\endDiscussion{}

\Discussion

\speaker{Dmitri Diakonov(NORDITA)}  The spin-$0$ diquark mass in 
vacuum has been measured in lattice computer simulations has been 
measured by the Bielefeld group (M.Hess {\it et al.}) and claimed to 
be around $700 \, MeV$. It seems that it is unbound.

\speaker{J.E. Horvath}  The $m_{D}$ appearing in the effective Lagrangian 
is not necessarily the vacuum mass to which calculations refer, but if the 
scalar diquark is indeed unbound the possibility of self-bound stars will be 
gone. Then diquark models may even be useful, although not for explaining 
small radii. 

\endDiscussion
 

\begin{thebibliography}{99}


\bibitem{benvenuto95} O. G. Benvenuto and G. Lugones, Phys. Rev D 51, 
1989 (1995); G. Lugones and O. G. Benvenuto, Phys. Rev D 52, 1276 
(1995) 

\bibitem{peng2000c61} G. X. Peng, H. C. Chiang, J. J. Yang, L. Li and B. 
Liu, Phys.  Rev.  C 61, 015201 (2000). 

\bibitem{horvath1} J. E. Horvath and J. A. de Freitas Pacheco, 
Int. J. of Mod. Phys. D 7, 19 (1998). 

\bibitem{li} X. -D. Li,  Z. -G. Lai and Z. -R. Wang,  Astro. Astrophys. 
303, L1 (1995) 

\bibitem{belyaev} V. M. Belyaev and Ya. I. Kogan, 
Phys. Lett. B 136, 273 (1984);  K. D.  Born, E. Laermann, N. Pirch, T. F. 
Walsh and P. M. Zerwas, 
Phys. Rev D 40, 1653 (1989). 

\bibitem{isgur} N. Isgur and J. Paton, Phys. Lett. B 124, 273 (1983); 
Phys. Rev D 31, 2910 (1985). 

\bibitem{vankerkwijk} M.H. van Kerkwijk, in {\it Proceedings of the 
ESO Workshop on Black Holes in Binaries and Galactic Nuclei},  eds. 
L. Kaper, E.M. van den Heuvel and P.A. Woudt (Springer-Verlag, 2001).

\bibitem{german} G. Lugones and J.E. Horvath, in preparation.

\bibitem{anselmino}  M. Anselmino, E. Pedrazzi, S. Elkin, S. Fredriksson and 
D.B. Litchemberg, Rev. Mod. Phys. 65, 1199 (1993).

\bibitem{CSW} M. Colpi, S. L.Shapiro and I. Wasserman, Phys. Rev. Lett. 
57, 2485 (1986).

\bibitem{pons} J.A. Pons, F.M. Walter, J.M. Lattimer, M. Prakash, 
R. Neuh\" auser and Penghui An, astro-ph/0107404 (2001).

\bibitem{MIT} O.G. Benvenuto and J.E. Horvath, MNRAS 241, 43 (1989).

\bibitem{ignazio} X-D. Li, I. Bombaci, M. Dey, J. Dey and E. P. J. van den Heuvel, 
Phys. Rev. Lett. 83, 3776 (1999).

\bibitem{foton} J.E. Horvath, Phys. Lett. B 242, 419 (1993). 

\end{thebibliography}
\end{document}